\documentclass[aps,preprint,prl]{revtex4}

\begin{document}

\newcommand{\be}{\begin{equation}}
\newcommand{\ee}{\end{equation}}
\newcommand{\ben}{\begin{eqnarray}}
\newcommand{\een}{\end{eqnarray}}
\newcommand{\nn}{\nonumber \\}
\newcommand{\ii}{\'{\i}}
\newcommand{\pp}{\prime}
\newcommand{\expq}{e_q}
\newcommand{\lnq}{\ln_q}
\newcommand{\quno}{q-1}
\newcommand{\qunoinv}{\frac{1}{q-1}}
\newcommand{\tr}{{\mathrm{Tr}}}

\draft


\title{Power-Law distributions and Fisher's information measure}

\author{F.~Pennini}
\thanks{E-mail:~pennini@fisica.unlp.edu.ar}

\author{A.~Plastino}
\thanks{E-mail:~plastino@fisica.unlp.edu.ar}

\address{Instituto de F\'{\i}sica La Plata (IFLP)\\
Universidad Nacional de La Plata (UNLP) and Argentine National
Research Council (CONICET)\\ C.C.~727, 1900 La Plata, Argentina}


\begin{abstract}
We show that thermodynamic uncertainties (TU) {\it preserve their
form} in passing from Boltzmann-Gibbs' statistics to Tsallis' one
provided that we express these TU in terms of the appropriate
variable conjugate to the temperature in a nonextensive context.
\vskip 2mm


\end{abstract}
\maketitle

\section{Introduction}
Thermodynamics' ``uncertainty'' (TU) relations  have been the
subject of much interesting work over the years (see, for
instance, \cite{ro,Mandelbrot,lavenda,Incerteza}). An excellent,
recent review is that of Uffink $\&$ van Lith \cite{Uffink}. A
possible starting point for the derivation of TU is statistical
inference \cite{roybook}. The pioneer work of Mandelbrot is in the
sense an obligatory reference \cite{Mandelbrot}. The interest in
TU  derives from the fact that for two pillars of $20$-$th$
century science,
  1) Heisenberg's uncertainty relations and 2) Bohr's complementarity principle, the
  existence of    a classical  analogue
has been suggested. Since such a proposal comes from none other
than Heisenberg and Bohr, the matter has been subjected to careful
scrutiny, specially in the case of a putative complementarity
between temperature and energy \cite{promine}. Although these
ideas have not received  general acceptance, several renowned
authors have defended them. We can cite, among others, Refs.
\cite{ro,Mandelbrot,lavenda}, whose claims remain still somewhat
controversial (see  \cite{Uffink,physto1,physto2}).

The purpose of the present effort is to make a (hopefully useful)
contribution to the ongoing discussion  by concentrating attention
on particular
   aspects of   Mandelbrot's  thermal
uncertainty derivation \cite{Mandelbrot}, that includes,  as an essential ingredient,
 the information measure introduced  by Fisher in the twenties
~\cite{roybook,f7}. Mandelbrot~\cite{Mandelbrot} was one of the first authors that linked
statistical physics with the theory of statistical inference, adopting the
viewpoint that one can work in statistical mechanics directly with probability
distributions over macroscopic variables, the phase space microscopic substructure
 being largely superfluous.

 Let $U$ stand for the internal energy.
Mandelbrot \cite{Mandelbrot} established  the {\it form} of the probability
distribution $p_\gamma (U)$ that allows for an appropriate description of the
energy  fluctuations of a system in contact with a heat bath at the
temperature $T=1/\gamma$. The ensuing distribution  turns out to be the
celebrated Gibbs' canonical one \cite{gibbs}, namely, an exponential
probability density $\exp{[-\gamma\,U]}$. A quite interesting uncertainty relation between mean
energy and inverse temperature can then be obtained, as detailed  below.
  The leading role in Mandelbrot's
 treatment is played by Fisher's information measure
 \cite{roybook,f7,pla2,pla4}

 \begin{equation}
I=\int \,d{\bf x}\,p_\theta({\bf x})\,\left\{\frac{\frac{\partial
p_\theta({\bf x})}{
\partial \theta}}{p_\theta({\bf x})}\right\}^2 =
\left\langle   \left[ \frac{1}{p_\theta} \frac{\partial p_\theta}{
\partial \theta} \right]^2 \right\rangle \label{ifisher}.
\end{equation}
 In order to get some insight into the significance of this measure consider
 a system that is specified by a physical
parameter $\theta$. Let {\bf x} be a stochastic variable  and
$p_\theta({\bf x})$ the probability density for this variable,
which depends on the parameter $\theta$.  An observer makes a
measurement of
 ${\bf x}$ and
has to best infer $\theta$ from this  measurement,
 calling the
  resulting estimate $\tilde \theta=\tilde \theta({\bf x})$. One
 wonders how well $\theta$ can be determined. Estimation theory
 asserts \cite{cramer} that the best possible estimator $\tilde
 \theta({\bf x})$, after a very large number of ${\bf x}$-samples
is examined, suffers a mean-square error $e^2$ from $\theta$ that
obeys a relationship involving Fisher's $I$, namely, $Ie^2=1$.
This ``best'' estimator is called the {\it efficient} estimator.
Any other estimator must have a larger mean-square error. The only
proviso to the above result is that all estimators be unbiased,
i.e., satisfy $ \langle \tilde \theta({\bf x}) \rangle=\,\theta
\label{unbias}$. Thus, Fisher's information measure has a lower
bound, in the sense that, no matter what parameter of the system
we choose to measure, $I$ has to be larger or equal than the
inverse of the mean-square error associated with  the concomitant
experiment. This result, i.e., \be \label{rao} I\,e^2\,\ge \,1,
 \ee
 is referred to as the
Cramer-Rao bound, and constitutes a very powerful statistical
result \cite{roybook}.

 {\it The central point of the present considerations is that,
 in addition to the  exponential distributions considered in \cite{Mandelbrot},
  we  often encounter power-law distributions (PLD) as well.}
 PLD  are certainly ubiquitous in physics (critical phenomena
are just  a conspicuous example \cite{goldenfeld}). It is well
known that in a statistical mechanics'
  context  power-law distributions arise quite naturally
  if the information measure one maximizes
 (subject to appropriate constraints)   in order to arrive at the equilibrium
 distribution  {\it is not Shannon's one} but a generalized one.  Much  effort in this
  respect has lately been reported. People  employ in  this type of extremizing
  processes  Tsallis' information measure as the quantity of interest
  (see \cite{gellmann,fromgibbs} and references therein).

   Taking into account  the importance of the concomitant  results \cite{gellmann},
    it is almost obligatory  to revisit the Fisher-Mandelbrot link
    by examining  non-exponential distributions of the power-law kind. A first
    step in this direction was taken in \cite{Incerteza}, where it was
    shown  that the  above
     mentioned ``exponential'' Fisher-Mandelbrot link
      cannot straightforwardly  be generalized to  power-law distributions so as to
      immediately yield thermal uncertainty relations.  Here we wish to explore into some more depth the issues
      investigated in \cite{Incerteza}
      by introducing the concept of ``effective'' energy into the pertinent discussion.

This paper is organized as follows. In Section II we introduce the
notion of ``effective energy'', the leitmotif of the present
considerations, and show that, with its help, a thermal
uncertainty relation can be derived. The meaning of this effective
energy is discussed in Section III and some conclusions are drawn
in Section IV.

\section{The effective energy and thermal uncertainty}
\label{THU} Escort distributions \cite{beck} are a typical feature
of Tsallis' thermostatistics and of its concomitant power-law
distributions \cite{TMP}. It is then necessary for our present
purposes to deal with the Fisher information notion as adapted to
a escort probability environment, i.e., with distributions of the
form  $P_\theta({\bf x})=p_\theta^q({\bf x})/\int d{\bf x}
p_\theta^q({\bf x})$. Following ~\cite{Incerteza,roybook,f7,Renyi}
we cast such a measure in the fashion

\be I=\int d{\bf x}\,P_\theta({\bf x})^{-1}\left[\frac{\partial
P_\theta({\bf x})}{\partial \theta }\right] ^{2}=\left\langle
\left[ \frac{1}{P_\theta}\frac{\partial P_\theta}{
\partial \theta} \right]^2 \right\rangle_{esc},
\label{Ifisher1}\ee where $p_\theta(\bf{x})$ is, again, the
probability density for the stochastic variable ${\bf x} \in
{\Re}^N$, $\theta$ a thermal parameter of the system, for example
the inverse temperature $\beta$, and $q$ a real parameter that can
be identified with Tsallis' nonextensivity index
\cite{t_jsp52,tsallisURL}. We speak then of ``Fisher measures in a
nonextensive context'' \cite {Renyi}. The associated Cramer-Rao
bound takes the form $I\, \Delta \theta\geq
q^2$~\cite{Incerteza,Renyi}.

We start our considerations by writing down the probability
distribution $P_{\beta}({\bf x})$ that extremizes
Tsallis' information measure~\cite{t_jsp52} subject to appropriate
constraints posed by our a priori knowledge. In a canonical thermodynamical system,
 the inverse temperature $\beta$ becomes a most appropriate parameter.
  In the present instance the piece of information supposedly
known a priori is  the generalized expectation value $U_q$ of the
internal energy $U({\bf x})$ \be U_q=\langle U \rangle_{esc}=\int
d{\bf x} P_{\beta}^q ({\bf x})\,U({\bf x}).\label{Uq} \ee

According to the Tsallis' formalism, as encapsulated by its {\it
optimal} Lagrange multipliers version \cite{OLM}  we  face the
following probability distribution
\begin{equation}
p_\beta({\bf x})  =\bar{Z}_q^{-1}\, e_q\left\{-\beta[U({\bf
x})-U_q ]\right\} ,\label{pd}
\end{equation}
 where $\beta$ is the  variational Lagrange
multiplier associated to $U_q$ and $\bar{Z}_q$ is the accompanying partition function (as
we sum over microstates no structure constant is needed~\cite{reif})
\begin{equation}
\bar{Z}_{q}=\int d{\bf x}\,e_q\left(-\beta\left(U({\bf x})-U_q \right)\right). \label{Zqp}
\end{equation}
We have made use of the so-called
generalized exponential \cite{tsallisURL}
\ben e_q(x)&=&[1+(1-q)x]^{\frac{1}{1-q}}
\,\,\,{\rm if}\,\,\,\,\,\,\,\,[1+(1-q)x] \ge 0 \nn &=& 0\,\,\,{\rm otherwise},
\een
a generalization of the exponential function, which is recovered when $q\rightarrow1$.

For the sake of an easier notation we shall try to omit hereafter,
as far as possible, writing down explicitly the variable ${\bf
x}$.  In order to accomplish  our purposes we need to evaluate the
integrand in (\ref{Ifisher1}). The definition of escort distribution is now given by
$\label{edb} P_{\beta}=  p_{\beta}^q/ \int d{\bf x}\,p_\beta^q.$

Returning to our present task and taking derivatives in
$P_\beta$ we find 

\be \frac{\partial P_\beta}{\partial \beta}=q\,P_\beta\left\{
p_\beta^{-1} \frac{\partial p_\beta}{\partial \beta}- \left\langle
p_\beta^{-1}\frac{\partial p_\beta}{\partial
\beta}\right\rangle_{esc} \right\},\label{P}
 \ee
  and proceed then to take
the pertinent derivatives  in  Eq.~(\ref{pd}), so as to confront
 \be p_\beta^{-1}\frac{\partial
p_\beta}{\partial \beta}=-\bar{Z}_q^{q-1} p_\beta^{q-1} (U -U_q),\label{p1}
 \ee
where we made  use of the fact  that $d\,\, ln \bar{Z}_q/d\,\,
\beta=0$, a result that one can derive by recourse to
 Eqs. (\ref{pd}) and (\ref{Zqp}) and/or can be encountered  in
~\cite{OLM}.

Notice that, from (\ref{pd}), we immediately obtain
\be \bar{Z}_q^{q-1} p_\beta^{q-1}=\left\{ 1-(1-q)\beta\left(U-U_q
\right)\right\}^{-1}. \label{Zq1} \ee At this precise stage we
define the quantity $E_{eff}$, an ``effective" energy  given by

\be E_{eff}=\frac{U_-U_q}{1-(1-q)\beta\left(U-U_q
\right)},\label{Eff} \ee which enables one to write

\be p_\beta^{-1}\frac{\partial p_\beta}{\partial \beta}= -
E_{eff}. \label{dlpb}\ee
Replacement of the  last two relations into
(\ref{P}) leads now to

\be \frac{\partial  P_\beta({\bf x})}{\partial  \beta}=
 q P_\beta({\bf x}) \left(- E_{eff} + \langle E_{eff}\rangle_{esc}\right), \ee
 and then to

 \be P_{\beta}^{-1}({\bf x})\left(\frac{\partial P_\beta({\bf x})}{\partial
\beta}\right)^2=q^2 P_\beta({\bf x}) \left(E_{eff}-\langle E_{eff}\rangle_{esc}\right)^2,
  \label{ll} \ee
which, when finally replaced into (\ref{Ifisher1}),  that is, integrating both sides of (\ref{ll})
over $d{\bf x}$,
 gives to Fisher's  information measure  the appearance

\begin{eqnarray}
I =q^{2} \left\langle \left(E_{eff}-\langle
E_{eff}\rangle_{esc}\right)^2 \right\rangle_{esc}.\label{IqH1}
\end{eqnarray}
A little additional algebra allows one finally to write

\be q^{-2}\,I  = \mu_{E_{eff}} \equiv \langle E_{eff}^2
\rangle_{esc}- \left\langle E_{eff}
\right\rangle_{esc}^2,\label{Iqf1} \ee so that the Cramer-Rao
bound gives
\be \label{effec} \mu_{E_{eff}}\, \Delta_{\beta} \ge
1, \ee
which is indeed an uncertainty relation.

\section{More on the  effective energy}
\label{EFF}

In this section we delve further into the   effective energy
concept. It is our aim here to show that $E_{eff}$ is the conjugate variable to $\beta$. For this purpose
we start by reminding the reader of the result \cite{OLM}
\be \label{OLM1} \int d{\bf x}\,p_\beta^q= \bar{Z}_q^{1-q},\ee
which together with the definition of escort distribution
 gives
 \ben \label{OLM2}  P_{\beta}=\bar{Z}_q^{q-1} p_\beta^q\een

 Thus,  it is possible to view the escort probabilities $P_\beta$ in a different
 light   by introducing  Eq. (\ref{Zq1}) into above and  obtaining
\be
P_\beta=\frac{p_\beta}{1-(1-q)\beta(U-U_q)}. \label{Pp}
\ee
   We thus establish a new connection between
   $P_\beta$ and $p_\beta$ in terms of the internal energy $U$. Further,  by introducing
   Eq. (\ref{Pp}) into the definition $\int d{\bf x}\,P_\beta (U-U_q)=0$ we also find that
\be \int d{\bf x}\, p_\beta \, E_{eff}=0,\label{E0} \ee which tells
us that $E_{eff}$ is a proper ``centered" variable. Also, by using
Eq. (\ref{Eff}) we can obtain $U-U_q$ as function of $E_{eff}$,
namely, \be \label{centro} U-U_q=\frac{E_{eff}}{1+(1-q)\beta
E_{eff}}, \ee so that replacing this into (\ref{pd}) we find

  \be  \label{invierte}
  p_\beta({\bf x})=
  \frac{1}{\bar Z_q}\,\,\frac{1}{e_{q}\left(\beta E_{eff}\right)}.
\ee

We point out now that
\begin{enumerate}
\item the probability (\ref{invierte})
 is to be compared to the Gibbs canonical distribution
\be  \label{GG}
  p_\beta({\bf x})=\frac{1}{Z_{Gibbs}}\,\,\frac{1}{e^{\beta U}}
  \ee
(in the limit $q \rightarrow 1$ (\ref{invierte}) tends to
(\ref{GG})), and
\item the $q$-form (Cf. Eq. (\ref{invierte})) of $ p_\beta({\bf x})$
neatly captures the fact that the probability distribution
{\it formally } depends, for fixed $q$, only upon $\beta$ and
$E_{eff}$.
\end{enumerate}

$\beta$ and $E_{eff}$ are thus our two ``conjugate" variables in
the present nonextensive instance. It does makes sense then  to
apply to them the Cramer-Rao bound. Remember also that Mandelbrot
never contemplated {\it actual} temperature fluctuations. He
assumed throughout that $T$ is fixed. Instead, the estimators are
random quantities \cite{Incerteza}. Such is the meaning one is to
read in a thermal uncertainty relation. And this meaning is here
precisely the same, via the somewhat artificial concept  of
effective energy.

\section{Conclusions}
\label{conclusions} We have shown that a thermal uncertainty
relation can be derived for power law distributions with the help
of the effective energy concept ($E_{eff}$), that, {\it as we have
discovered here}, is the conjugate variable to the inverse
temperature in a nonextensive Tsallis setting.

The concept of thermal uncertainty applies to conjugate variables.
These are
\begin{enumerate}
\item $\beta$ and $U$ for $q=1$, and
\item $\beta$ and $E_{eff}$ for $q \ne 1$. Of course,  $E_{eff} \rightarrow U$
in the limit $q \rightarrow 1$.
\end{enumerate}

Summing up, if one looks for the appropriate conjugate variables,
the concept of thermal uncertainty continues to make sense in a
nonextensive setting, contrary to what was stated in
\cite{Incerteza}, where an uncertainty relation was looked for
between $\beta$ and $U$, and, of course, could not be found.

\section{Acknowledgements}  The authors are indebted to Prof. C. Tsallis for
very helpful discussions. Also, F.~ Pennini acknowledges financial
support from CONICET, Argentina.


\begin{references}

\bibitem{ro} L.~ Rosenfeld, in {\it Ergodic theories}, edited by P.~Caldirola (Academic Press, NY, 1961).

\bibitem{Mandelbrot} B.~Mandelbrot, {\it Ann. Math. Stat.} {\bf
33} (1962) 1021; {\it IRE Trans. Inform. Theory}, {\bf IT-2}
(1956) 190; {\it J.~ Math.~ Phys.} {\bf 5} (1964) 164.

\bibitem{lavenda} B.~ Lavenda, {\it Int. J. Theor. Phys.} {\bf 26}
(1987) 1069; {\bf 27} (1988) 451; {\it J. Phys. Chem. Sol.} {\bf 49}
(1988) 685; {\it Statistical physics: a probabilistic approach}
(J. Wiley, NY, 1991).

\bibitem{Incerteza} F.~ Pennini, A.~ Plastino, A.R.~ Plastino and M.~ Casas,
{\it Physics Letters A} {\bf 302} (2002) 156.

\bibitem{Uffink} J.~ Uffink and J.~van Lith, {\it Foundations of
Physics} {\bf 29} (1999) 655.


\bibitem{roybook}  B.R.~ Frieden, {\it Physics from Fisher information}
(Cambridge University Press, Cambridge, England, 1998).


\bibitem{promine} N.~ Bohr, {\it Collected works}, edited by J.
Kalckar (North-Holland, Amsterdam, 1985), Vol. 6, pp. 316-330 and
376-377; A. Pais, {\it Niels Bohr's times in physics, philosophy,
and polity} (Clarendon Press, Oxford, 1991).

\bibitem{physto1}  C.~ Kittel, {\it Phys. Today}
(May 1988) 93.

\bibitem{physto2}  B.B.~ Mandelbrot, {\it Phys. Today}
(January 1989) 71.

\bibitem{f7}  B.R.~ Frieden and B.H.~ Soffer, {\it Phys. Rev. E} {\bf 52} (1995) 2274.

\bibitem{gibbs} J.W.~ Gibbs, {\it Elementary principles in
statistical mechanics} (Yale University Press, 1903).


\bibitem{pla2}  A.R.~ Plastino and A.~Plastino, {\it Phys. Rev. E} {\bf 54} (1996) 4423.

\bibitem{pla4}  A.~ Plastino, A.R.~ Plastino, and H.G.~ Miller, {\it Phys.
Lett. A} {\bf 235} (1997) 129.

\bibitem{cramer} H.~ Cramer, {\it Mathematical methods of statistics},
(Princeton University Press, Princeton, NJ, 1946).

\bibitem{goldenfeld} N.~ Goldenfeld, {\it Lectures on phase transitions and the
renormalization group} (Addison-Wesley, NY, 1992).

\bibitem{gellmann} M.~Gell-Mann and C.~ Tsallis, ``Nonextensive Entropy -
Interdisciplinary Applications'',  (Oxford University Press,
Oxford, 2003).


\bibitem{fromgibbs} A.R.~ Plastino, A.~ Plastino, {\it Phys. Lett.
A} {\bf 193} (1994) 251.


\bibitem{beck}  C.~ Beck and F.~ Schl\"{o}gl, {\it Thermodynamics of chaotic
systems} (Cambridge University Press, Cambridge, England, 1993).

\bibitem{TMP}
C.~Tsallis, R.S.~Mendes and A.R.~Plastino, {\it Physica A} {\bf
261} (1998) 534.


\bibitem{Renyi}  F.~ Pennini, A.R.~ Plastino and A.~ Plastino, {\it Physica A}
{\bf 258} (1998) 446.

\bibitem{t_jsp52} C.~Tsallis, {\it J.~Stat.~Phys.} {\bf 52} (1988) 479.


\bibitem{tsallisURL}
A periodically updated bibliography on nonextensive
thermostatistics can be found in the URL
http://tsallis.cat.cbpf.br/biblio.htm

\bibitem{OLM}
S.~Mart\'{\i}nez, F.~Nicol\'{a}s, F.~Pennini and A.~Plastino,
{\it Physica A} {\bf 286} (2000) 489.





\bibitem{reif} F.~ Reif, {\it Statistical and thermal physics}
(McGraw-Hill, NY, 1965).



\end{references}
 \end{document}